\documentstyle[aps,multicol,graphics,prb]{revtex}
\newcommand{\bfr}{{\bf r}}
\newcommand{\bfx}{{\bf x}}
\newcommand{\bfy}{{\bf y}}
\newcommand{\bfD}{{\bf \Delta}}
\newcommand{\bfJ}{{\bf J}}
\begin{document}
\draft
\title{Current-voltage characteristics of the two-dimensional $XY$ model
with Monte Carlo dynamics}
\author {Beom Jun Kim}
\address {Department of Theoretical Physics,
Ume{\aa} University,
901 87 Ume{\aa}, Sweden}
\preprint{\today}
\maketitle
\thispagestyle{empty}
\begin{abstract}
Current-voltage characteristics and the linear resistance of the two-dimensional 
$XY$ model with and without external uniform current driving are studied by 
Monte Carlo simulations.
We apply the standard finite-size scaling analysis 
to get the dynamic critical exponent $z$ at various temperatures. 
From the comparison with the resistively-shunted 
junction dynamics, it is concluded that $z$ is universal 
in the sense that it does not depend on details of dynamics.
This comparison also leads to the quantification of the time in the Monte 
Carlo dynamic simulation.
\end{abstract}


\begin{multicols}{2}

Equilibrium properties of the two-dimensional (2D) $XY$ model
have been studied extensively and there are well-established consensus
on them.~\cite{review}
Dynamic properties of 2D $XY$ model, however, are still
under strong debate: For example, some studies on 
the dynamic critical exponent $z$ seem to contradict each 
other.~\cite{bjkim99a,lars,shortXY,pierson,simkin97,holmlund,weber96}
Investigation of dynamic properties
requires a set of dynamic equations or impositions of update rules
and there exist various possible choices for the $XY$ model. On the one hand,
there are Langevin-type stochastic equations like
the resistively shunted junction (RSJ) dynamics for $XY$ model with
phase representation,~\cite{bjkim99a,lars,simkin97}  and the relaxational 
dynamics (often called the time-dependent Ginzburg Landau dynamics) 
for both phase~\cite{bjkim99a,lars,jonsson97} and vortex 
representations.~\cite{holmlund}
On the other hand, it is also possible to use Monte Carlo (MC) simulations
to study dynamic behaviors~\cite{weber96,jrlee} although the
identification of a Monte Carlo update step with a time step still
lacks a complete rigorous justification.~\cite{mychoi:mc}
An understanding of the dynamic transport properties, like   
the current-voltage characteristics, in the  
above dynamic models are 
important also in the practical sense, since such properties are
directly measured in experiments on $XY$-like realizations such as 
Josephson-junction arrays and high-$T_c$ superconductors.

We here use the standard MC simulation method to study the current-voltage 
characteristics of the $XY$ model with phase representation.
The purpose of the current investigation is to find the dynamic
critical exponent $z$ in MC dynamic simulation and compare it
with existing results from other dynamics. It then turns out that
it is possible to quantify the MC time scales from the comparison
with RSJ dynamics (see Ref.~\onlinecite{nowak} for a recent
discussion on the time quantification of MC dynamics). This opens
a possibility to use MC simulation, which is usually much more efficient
than direct numerical integrations of stochastic equations of motion,
to examine many interesting long-time dynamic properties of the $XY$ models.

The fluctuating twist boundary condition (FTBC) has been introduced first in
the static MC simulation,~\cite{olsson}  and
later has been extended to RSJ dynamics.~\cite{bjkim99a} 
In the computations of dynamic quantities, it has been shown that the 
FTBC has some advantage over the periodic boundary condition: 
For example, the  fluctuating linear resistance can be calculated
in a much more convenient way in the FTBC.~\cite{bjkim99a,lars}
We start from the RSJ dynamic equations of motion for the 2D $XY$ model
on a square lattice in the presence
of an external current in the $x$ direction with the current density 
$\bfJ = (J,0)$ under the FTBC.~\cite{bjkim99a}
The phase variable $\theta_i$ on the $i$th site at position $\bfr_i$ satisfies~\cite{comment:unit}
\begin{eqnarray} \label{eq:theta}
\dot\theta_i &=& 
-\sum_j G_{ij}{\sum_k}^{'} [ \sin(\theta_j - \theta_k - 
{\bfD}\cdot {\bf r}_{jk}) + \eta_{jk}]  \nonumber \\
&\equiv& h_i 
- \sum_j G_{ij}{\sum_k}^{'}\eta_{jk}  ,
\end{eqnarray}
where $G_{ij}$ is the lattice Green function, the primed
summation is over nearest-neighbor sites ($k$) of  $j$, 
${\bfD} = (\Delta_x, \Delta_y)$ is the fluctuating twist variable, and 
$\bfr_{jk} \equiv \bfr_k - \bfr_j$. The thermal noise $\eta_{ij}$ is
white noise satisfying $\langle \eta_{ij} \rangle = 0$ and 
$ \langle \eta_{ij}(t) \eta_{kl} (0)\rangle =
2T(\delta_{ik}\delta_{jl} -  \delta_{il}\delta_{jk})\delta(t) ,
$
with the temperature $T$ in units of Josephson coupling strength. 
The equations of motion for $\bfD$ are written as
\begin{eqnarray}
\frac{d{\Delta}_x}{dt} &=& \frac{1}{L^2} \sum_{\langle ij\rangle_x}
   \sin(\theta_i - \theta_j - \Delta_x) +\eta_{\Delta_x} - J  \nonumber \\
   &\equiv& h_x + \eta_{\Delta_x} - J   \label{eq:dot:deltax} , \\
\frac{d {\Delta}_y }{dt} &=& \frac{1}{L^2} \sum_{\langle ij\rangle_y}
    \sin(\theta_i - \theta_j - \Delta_y) +\eta_{\Delta_y}  \nonumber \\
 &\equiv& h_y + \eta_{\Delta_y}    \label{eq:dot:deltay}  ,
\end{eqnarray}
where $\sum_{\langle ij\rangle_x}$ denotes the summation over all
links in the $x$ direction, and the thermal noise terms satisfy
$\langle \eta_{\Delta_x} \rangle  = \langle \eta_{\Delta_y} \rangle 
= \langle \eta_{\Delta_x} \eta_{\Delta_y}  \rangle = 0$ and 
$\langle \eta_{\Delta_x}(t) \eta_{\Delta_x} (0)  \rangle =
\langle \eta_{\Delta_y}(t) \eta_{\Delta_y} (0)  \rangle =
(2T/L^2) \delta (t)$.
The above Langevin-type equations of motion 
(\ref{eq:theta})-(\ref{eq:dot:deltay})
can be cast into the Fokker-Planck equation for the probability 
distribution function $P$:
\begin{eqnarray}
\frac{\partial P}{\partial t} = & &
- \sum_i \frac{\partial}{\partial \theta_i} h_i P
- \frac{\partial}{\partial \Delta_x} h_x P
- \frac{\partial}{\partial \Delta_y} h_y P \nonumber \\
& & + T\sum_{ij} G_{ij} \frac{\partial^2 P}{\partial \theta_i \partial \theta_j}
+ \frac{T}{L^2}\left( \frac{\partial^2 P}{\partial \Delta_x^2}
                    +  \frac{\partial^2 P}{\partial \Delta_y^2} \right), 
\end{eqnarray}
whose stationary solution ($\partial P/\partial t = 0$)
in the form $P = e^{-H/T}$ consequently leads to the 
effective Hamiltonian:~\cite{comment:H}
\begin{equation} \label{eq:H}
H = -\sum_{\langle i j\rangle} \cos(\theta_i - \theta_j - \bfD\cdot\bfr_{ij})
   + L^2 \bfJ\cdot\bfD , 
\end{equation}
where the summation runs over all the nearest-neighboring bonds constituting the array.
The Hamiltonian (\ref{eq:H}) is not invariant under
the transformation $\Delta_x \rightarrow \Delta_x + 2\pi$, 
in contrast to the original RSJ equations. However, we note
that in MC simulations the energy difference of configurations, 
not the energy itself,  determines the time evolution of the system.
Thus a direct use of Eq.~(\ref{eq:H}) in the MC simulations does not cause any 
problem.

In the MC simulations, we first pick one site and
try to rotate the phase angle at this site by an amount randomly
chosen in $[-\delta\theta, \delta\theta]$ (we call $\delta\theta$ 
the trial angle range). This MC try  
is accepted according to the standard Metropolis
algorithm applied to the Hamiltonian (\ref{eq:H}). 
After sweeping through all the lattice sites to update
all the phase variables, we update the fluctuating twist variables $\Delta_x$
and $\Delta_y$ in a similar way: we try to rotate $L\Delta_x$ and
$L\Delta_y$
within the angle range $\delta\Delta$.
(For convenience, we use $\delta\theta = \delta\Delta$.)
This complete update of the phase variables and the twist
variables  constitute one MC step. It is this MC step which is
identified with one time step.
A static quantity in equilibrium does not depend on the choice of 
$\delta\theta$; 
the choice of $\delta\theta$ only affects the convergence rate of the quantity (too small
and too large $\delta\theta$ worsen the convergence).
This is contrary to a  dynamic quantity which in general is expected to depend on  $\delta\theta$.  
In Ref.~\onlinecite{nowak}, from the study of the anisotropic magnetic
particle in an external magnetic field, it has been shown that as far
as long-time behaviors are
concerned the real dynamics
and MC dynamics become in good agreements when the time evolution
in MC simulation becomes slow enough. If such a result also did carry
over to our model it would for our MC simulation imply
that the MC dynamics
give the same current-voltage characteristics as the RSJ dynamics after
a suitable normalization of time, when $\delta\theta$ is sufficiently small. 
As will be shown below, we find that this is indeed the case.

\begin{figure}
\centering{\resizebox*{!}{6.5cm}{\includegraphics{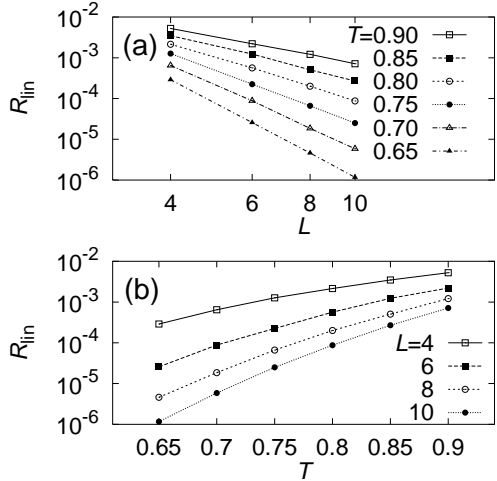}}}
\caption{The linear resistance $R_{\rm lin}$ calculated at the trial angle
range $\delta\theta=\pi/6$ in the absence
of external current (a) vs system size $L$ and (b) vs temperature $T$.
In (a), the dynamic critical exponent $z$ defined as $R_{\rm lin} \sim L^{-z}$
is shown to increase as $T$ is decreased (see Fig.~\ref{fig:zRlin} for $z$ at
various $\delta\theta$). 
}
\label{fig:Rlin}
\end{figure}

\begin{figure}
\centering{\resizebox*{!}{6.0cm}{\includegraphics{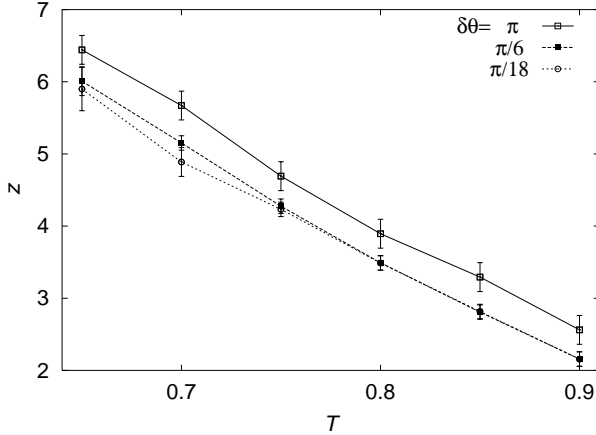}}}
\caption{The dynamic critical exponent $z$ from 
the linear resistance $R_{\rm lin}$ for the trial angle
range $\delta\theta=\pi$, $\pi/6$, and $\pi/18$ (from top to bottom).
As $\delta\theta$ is decreased, $z$ is shown to saturate and
$z(\delta\theta = \pi/6$) coincides well with $z(\delta\theta=\pi/18$).
}
\label{fig:zRlin}
\end{figure}

In the absence of external current [$\bfJ = 0$ in Eq.~(\ref{eq:H})], 
we first calculate $L\Delta_x$ as a function of time, 
typically up to $t=10^9$-$10^{10}$ after equilibration, and then compute 
the fluctuating linear resistance~\cite{bjkim99a}
\begin{equation} 
\label{eq:Rlin}
R_{\rm lin} =  \frac{1}{2 T } \frac{1}{\Theta} \langle [ L\Delta_x(\Theta) - L\Delta_x(0) ]^2 \rangle ,
\end{equation} 
where $\Theta$ is chosen to be sufficiently large and 
$\langle \cdots \rangle$ is substituted by the time average 
(see Refs.~\onlinecite{lars} and \onlinecite{bjkim00}).
The linear resistance is closely related with the time scale $\tau$,  
defined by the time spent between the jumps of $L\Delta_x$ by $2\pi$ 
(see Fig.~8 in Ref.~\onlinecite{bjkim99a}).
\begin{equation} \label{eq:Rtau} 
R_{\rm lin} =  \frac{1}{2T}\frac{ (2\pi)^2 }{\tau} .
\end{equation}
In other words, the linear resistance is obtained from the 
equilibrium fluctuations of $L\Delta_x$ which has a time scale $\tau$
related with jumps by $2\pi$.
Since the computation of $R_{\rm lin}$ in practice is more straightforward
than $\tau$, we calculate the former and $\tau$ is simply obtained 
by Eq.~(\ref{eq:Rlin}).~\cite{comment:Rtau}
As is well known, the whole low-temperature phase of the 2D $XY$ model 
is quasi-critical and the dynamic critical exponent $z$ is 
defined by $\tau \sim L^z$ due to the infinite correlation length.
Alternatively, $z$ can be computed from Eq.~(\ref{eq:Rtau})
\begin{equation} \label{eq:RlinL}
R_{\rm lin} \sim L^{-z} 
\end{equation}
at a given temperature $T$.
Although there exist several studies with a different conclusion,~\cite{comment:z} 
there is nevertheless a growing consensus that the 2D $XY$ model in the low-temperature phase 
has a $z$ which is a monotonically decreasing function of $T$.
Figure~\ref{fig:Rlin} displays (a) $R_{\rm lin}$ vs $L$ at various $T$ 
and (b) $R_{\rm lin}$ vs $T$ for various $L$, for the trial angle 
range $\delta\theta = \pi/6$. As expected, the dynamic critical exponent $z$
defined by Eq.~(\ref{eq:RlinL}) becomes larger as $T$ is lowered.
The dependence of $R_{\rm lin}$ on $\delta\theta$ is also studied: we find
that although the value of $R_{\rm lin}$ strongly depends on $\delta\theta$,
the exponent $z$ appears to saturate as $\delta\theta$ is decreased, as shown
in Fig.~\ref{fig:zRlin}, where $z(\delta\theta = \pi/18) \approx z(\pi/6)$
(the bigger deviations at lower temperatures are due to the poor
convergences in $R_{\rm lin}$ at $\delta\theta=\pi/18$ since very long
time evolution is needed due to very large characteristic time).
Since the time scale is inversely proportional to the linear resistance
[see Eq.~(\ref{eq:Rtau})], Fig.~\ref{fig:zRlin} implies that
the time scales at $\delta\theta = \pi/6$ and $\pi/18$ are simply
proportional to each other in a broad range of temperatures and
system sizes; we numerically find
$\tau(T,L,\delta\theta=\pi/6) \approx c\tau(T,L,\delta\theta=\pi/18)$ 
with $c \approx 0.14$ independent of $T$ and $L$.

\begin{figure}
\centering{\resizebox*{!}{6.0cm}{\includegraphics{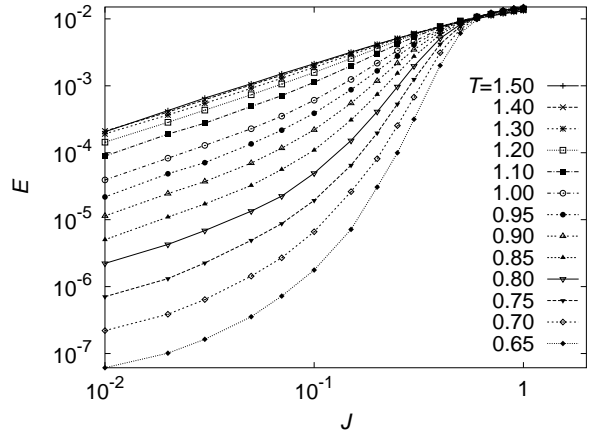}}}
\caption{Current-voltage characteristics (electric field $E$ vs 
current density $J$) for $L=8$ and the trial angle range $\delta\theta=\pi/6$
at various temperatures. As $J$ is decreased the current-voltage
characteristics become Ohmic ($E \propto J$) due to finite-size effects.
}
\label{fig:IV}
\end{figure}

If driven by an external current in the $x$ direction, 
the twist variable $\Delta_x$ decreases as time goes on 
since the smaller $\Delta_x$ gives the lower energy [see Eq.~(\ref{eq:H})],
and the system develops a voltage drop across the
sample with the electric field 
\begin{equation} \label{eq:E}
E  = -\langle \dot\Delta_x  \rangle.
\end{equation} 
We perform simulations for $L=4$, 6, 8, and 10  
at $0.65 \leq T \leq 1.5$ with $\delta\theta = \pi/6$ and $\pi/18$:
Figure~\ref{fig:IV}, as an example, displays the current-voltage characteristics ($E$ vs $J$) 
at various temperatures for $L=8$ and $\delta\theta = \pi/6$.
At low currents, the current-voltage characteristics become linear ($E \sim J$)
at any temperatures in accordance with Ref.~\onlinecite{katya} for RSJ dynamics.

\begin{figure}
\centering{\resizebox*{!}{6.0cm}{\includegraphics{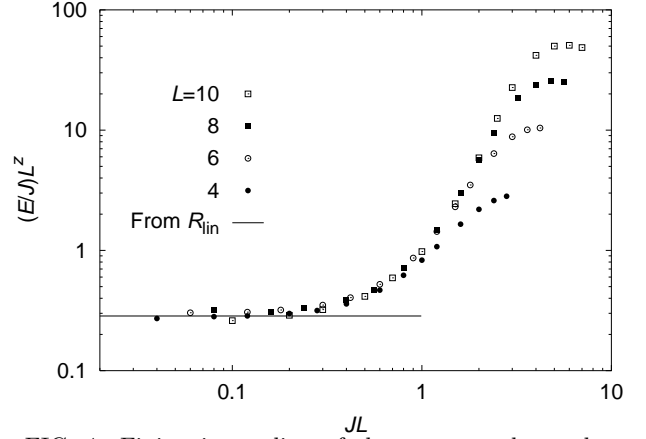}}}
\caption{Finite-size scaling of the current-voltage characteristics,
$(E/J)L^z$ vs $JL$, at $T=0.80$ and $\delta\theta =\pi/6$.
The dynamic critical exponent $z$, the only one free parameter in
the scaling plot, is found to $z=3.5(1)$ at $T=0.80$, in agreement
with the value in Fig.~\ref{fig:zRlin} obtained in the absence
of external driving.
The horizontal full line denotes $R_{\rm lin}L^z$ with $z=3.5$ 
obtained in the absence of an external current, confirming that the simulation 
results with and without external currents are fully consistent to each other.
}
\label{fig:IVscale}
\end{figure}

In the low-temperature phase of the 2D $XY$ model,
the dynamic scaling in Ref.~\onlinecite{FFH} for the current-voltage
characteristics takes the form~\cite{bjkim99a,katya} 
\begin{equation} \label{eq:IVscale}
E/J = L^{-z}f(JL), 
\end{equation}
where $z$ and the scaling function $f(x)$ depend on $T$.
In Fig.~\ref{fig:IVscale}, the finite-size scaling is
exhibited in the form $(E/J)L^z = f(JL)$ 
at $T=0.80$ for $\delta\theta = \pi/6$. In this construction 
$z$ is the only one free parameter and $z=3.5(1)$ is obtained. 
The solid line in Fig.~\ref{fig:IVscale}
is obtained from $R_{\rm lin}$ computed above in the absence of
external currents.

Table~\ref{tab:z} summarizes our MC results for $z$ 
with and without external currents,
i.e., $z$ from $R_{\rm lin}$ without current
driving in Fig.~\ref{fig:zRlin}, and $z$ 
from the finite-size scaling of the current-voltage characteristics 
in the presence of external current (see
Fig.~\ref{fig:IVscale}). As seen from the table, these two different
determinations are in agreement. 
In Table~\ref{tab:z} we have also included the
$z$ values obtained for the     
RSJ dynamics in Ref.~\onlinecite{bjkim99a},
and again find agreement. From this we conclude that 
the $z$ values from the Monte Carlo dynamic
simulation in the present work agree with the corresponding values 
from the RSJ dynamics. 

\begin{figure}
\centering{\resizebox*{!}{6.0cm}{\includegraphics{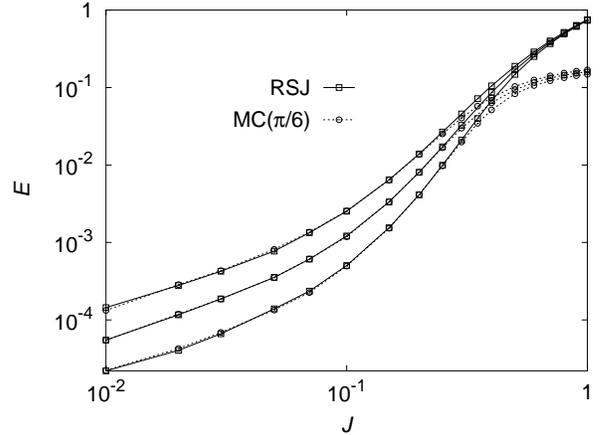}}}
\caption{Comparison of the current-voltage characteristics ($E$-$J$)
of the MC simulation (from Fig.~\ref{fig:IV}) and the RSJ dynamics 
(in Ref.~\protect\onlinecite{katya}) for $L=8$ at $T=0.90$, 0.85, and 0.80
(from top to bottom): 
Multiplications by factors 11.6, 10.9, and 10.1 for $T=0.90$, 0.85, and
0.80, respectively, to $E$ from MC simulations make two curves coincide 
very well in a broad range of external current density. 
}
\label{fig:quant}
\end{figure}

The above agreements suggest that some features of the real time
dynamics for the $XY$ model can be studied by MC simulations. The
practical advantage is that MC simulations are 
usually much more efficient than numerical integrations of stochastic
dynamic equations like RSJ equations. 
In Fig.~\ref{fig:quant} we carry the comparison one step further: we
directly compare the current-voltage
characteristics obtained in this paper from MC at $\delta\theta=\pi/6$ with that in 
Ref.~\onlinecite{katya} from RSJ, for $L=8$ at $T=0.90$, 0.85, and 0.80 (from
top to bottom). 
It is shown that multiplications of $E$ from MC by factors 11.6, 10.9,
and 10.1 for $T=0.90$, 0.85, and 0.80, respectively, 
result in perfect agreements within a broad range of $J$
(the temperature-dependence of the conversion factor has also been
found in Ref.~\onlinecite{nowak}).
Since the resistance is inversely proportional to the time scale 
[see Eq.~(\ref{eq:Rtau})], Fig.~\ref{fig:quant} leads to the result
that the time scale of the MC simulation at $\delta\theta=\pi/6$ 
for a given temperature is larger than RSJ dynamic simulation by the 
above constant factor.

In summary, we have performed the MC dynamic simulation
of the 2D $XY$ model subject to the fluctuating twist
boundary condition with and without external currents.
Through the use of the standard finite-size scaling method,
the dynamic critical exponent $z$ has been obtained as a function of
temperature, and found to be in accordance with RSJ dynamic results
in Ref.~\onlinecite{bjkim99a}, as well as with the scaling prediction
by Minnhagen {\it et al}. in 
Ref.~\onlinecite{minnhagen}. 
It has also been shown that the current-voltage characteristics 
obtained from MC dynamic simulation is in a good agreement
with that from the RSJ dynamics, leading to the quantification
of time in MC simulation.~\cite{foot} This implies that it is possible to use
MC dynamic simulation, which is much more efficient than numerical
integration of stochastic dynamic equations in many cases, to study
long-time behaviors of the real dynamics of the $XY$ model.
For example, the time-quantified MC simulation can be useful 
to study glass-like dynamic behaviors of the $XY$ models 
with disorder,~\cite{sjlee}
as well as to investigate current-voltage characteristics
of the $XY$ model in very low temperatures where RSJ simulation
takes too much time to get converged results.~\cite{katya2}

The author is grateful to Petter Minnhagen for useful discussions.
This work was supported by the Swedish Natural Research Council
through Contract No. FU 04040-332.


\begin{thebibliography}{10}

\bibitem{review}
See for reviews, P. Minnhagen, Rev. Mod. Phys. {\bf 59},  1001  (1987);
{\em Proceedings of the 2nd CTP workshop on Statistical Physics: KT Transition
  and Superconducting Arrays}, edited by D. Kim, J.~S. Chung, and M.~Y. Choi
  (Min-Eum Sa, Seoul, 1996), and references therein.

\bibitem{bjkim99a}
B.~J. Kim, P. Minnhagen, and P. Olsson, Phys. Rev. B {\bf 59},  11506  (1999).

\bibitem{lars}
L.~M. Jensen, B.~J. Kim, and P. Minnhagen, Europhys. Lett. 
{\bf 49},  644 (2000); Phys. Rev. B {\bf 61},  15412 (2000).

\bibitem{shortXY}
H.~J. Luo and B. Zheng, Mod. Phys. Lett. B {\bf 11},  615  (1997);
A.~J. Bray, A.~J. Briant, and D.~K. Jervis, Phys. Rev. Lett. {\bf 84},  1503
  (2000).

\bibitem{pierson}
S. M. Ammirata, M. Friesen, S. W. Pierson, L. A. Gorham, J. C. Hunnicutt,
M. L. Trawick, C. D. Keener, Physica C {\bf 313},  225  (1999);
S. W. Pierson, M. Friesen, S. M. Ammirata, J. C. Hunnicutt,
and L. A. Gorham, Phys. Rev. B {\bf 60},  1309  (1999);
S.~W. Pierson and M. Friesen, Physica B {\bf 284},  610  (2000).

\bibitem{simkin97}
M.~V. Simkin and J.~M. Kosterlitz, Phys. Rev. B {\bf 55},  11646  (1997).

\bibitem{holmlund}
K. Holmlund and P. Minnhagen, Phys. Rev. B {\bf 54},  523  (1996);
Physica C {\bf 292},  255  (1997).

\bibitem{weber96}
H. Weber, M. Wallin, and H. Jensen, Phys. Rev. B {\bf 53},  8566  (1996).

\bibitem{jonsson97}
A. Jonsson and P. Minnhagen, Phys. Rev. B {\bf 55},  9035  (1997).

\bibitem{jrlee}
J.-R. Lee and S. Teitel, Phys. Rev. B {\bf 50},  3149  (1994).

\bibitem{mychoi:mc}
M. Y. Choi and B. A. Huberman, Phys. Rev. B {\bf 28}, 2547 (1983); {\bf 29} 2796 (1984).

\bibitem{nowak}
U. Nowak, R.~W. Chantrell, and E.~C. Kennedy, Phys. Rev. Lett. {\bf 84},  163
(2000); R. Smimov-Rueda, O. Chubykalo, U. Nowak, R. W. Chantrell,
and J. M. Gonzalez, J. Appl. Phys. {\bf 87},  4798  (2000).

\bibitem{olsson}
P. Olsson, Phys. Rev. B {\bf 46},  14598  (1992); {\bf 52},  4511  (1995);
{\bf 52},  4526  (1995).

\bibitem{comment:unit}
In RSJ model, we measure the time, the current, the energy, and the temperature
  in units of $\hbar/2eRI_c$, $I_c$, $\hbar I_c/2e$, and $\hbar I_c/2ek_B$,
  respectively ($I_c$ is the single junction critical current and $R$ is the
  shunt resistance).

\bibitem{comment:H}
Alternatively, Eq.~(\ref{eq:H}) also can be obtained from
the effective Hamiltonian for the free boundary condition
in the presence of an external current 
[see M.~Y. Choi, Phys. Rev. B {\bf 46},  564  (1992)] 
through a simple change of variables (see Ref.~\onlinecite{bjkim99a})
and then imposing the periodicity,
$\theta_{i + L{\hat \bfx}} = \theta_{i + L{\hat \bfy}}  = \theta_i$.

\bibitem{bjkim00}
B.~J. Kim, Phys. Rev. B {\bf 62},  644  (2000).

\bibitem{comment:Rtau}
Equation~(\ref{eq:Rtau}) was verified numerically when the system size was
  small in this work. We also confirmed it from data presented in
  Ref.~\onlinecite{bjkim99a} for the resistively-shunted junction dynamics.

\bibitem{comment:z}
The short-time relaxation method [see Z.~B. Li, L. Sch{\"u}lke, and 
B. Zheng, Phys. Rev. Lett. {\bf 74},  3396 (1995)]
applied for 2D $XY$ model
  with Monte Carlo dynamic simulation has found $z\approx 2$ in the whole
  low-temperature phase [see, e.g., Ref.~\onlinecite{shortXY}], while the
  scaling method by Pierson {\it et al.} for various experiments and
  simulations has lead to a relatively large but temperature-independent value
  $z \approx 6$ [Ref.~\protect\onlinecite{pierson}].

\bibitem{katya}
K. Medvedyeva, B.~J. Kim, and P. Minnhagen, Phys. Rev. B (in press).

\bibitem{FFH}
D.~S. Fisher, M.~P.~A. Fisher, and D.~A. Huse, Phys. Rev. B {\bf 43},  130
  (1991).

\bibitem{minnhagen} P. Minnhagen, O. Westman, A. Jonsson, and
  P. Olsson, Phys. Rev. Lett. {\bf 74}, 3672 (1995).

\bibitem{foot} With material parameters $I_c = 7$mA and
the shunt resistance $R = 2$m$\Omega$, one MC step in
this paper is found to approximately corresponds to 
$10^{-10}$sec.

\bibitem{sjlee} S.~J. Lee (private communication).

\bibitem{katya2} K. Medvedyeva, B.~J. Kim, and P. Minnhagen, unpublished.


\end{thebibliography}

\narrowtext

\begin{table}
\caption{ Dynamic critical exponent $z$ of 2D $XY$ model
subject to Monte Carlo dynamics. $z(R_{\rm lin})$ is from
the scaling of the linear resistance at $\delta\theta=\pi/6$
(Fig.~\ref{fig:zRlin}) and
$z(E$-$J)$ is from the finite-size scaling of the current-voltage
characteristics (see Fig.~\ref{fig:IVscale}).  The results from the RSJ dynamics in 
Ref.~\protect\onlinecite{bjkim99a} are also presented for comparisons.
}

\begin{tabular}{l c c c c c c}
$T$              & 0.65   & 0.70   & 0.75   & 0.80 & 0.85 & 0.90  \\ \hline
$z(R_{\rm lin}) $& 6.0(2) & 5.2(1)    & 4.3(1)    & 3.5(1)  & 2.8(1)  & 2.2(1)   \\ \hline
$z(E$-$J)$       & 5.9(3) & 5.0(2) & 4.3(2) & 3.5(1)  & 2.8(1)  & 2.2(1)  \\ \hline
$z({\rm RSJ})^{\rm a}$  &  -      &   -    &   -    &   3.3(1)  & 2.7(1) & 2.0(1) 
\end{tabular}
$^{\rm a}$ Reference~\onlinecite{bjkim99a}
\label{tab:z}
\end{table}

\end{multicols}
\end{document}